\documentclass[aps,prl,preprint,showpacs,superscriptaddress]{revtex4}

\newcommand{\ts}{\textsuperscript}

\usepackage[dvips]{graphicx}

\usepackage{xspace}
\newcommand{\etal}{\textit{et al.}\xspace}
\newcommand{\neETwoPlus}{722(9) keV\xspace}  
\newcommand{\ioi}{``Island of Inversion''\xspace}
\newcommand{\etwop}{\ensuremath{E(2^+_1)}\xspace}
\newcommand{\beupl}{\ensuremath{B(E2;~0^+_{\mathrm{gs}}\rightarrow2^+_1)}\xspace} 
\newcommand{\gammaray}{\ensuremath{\gamma}-ray\xspace}
\newcommand{\kn}[1]{}   

\begin{document}

  %
  %
  \title{Spectroscopy of \ts{32}Ne and the \ioi}

  \newcommand{\ariken}{  \affiliation{RIKEN Nishina Center, Wako, Saitama 351-0198, Japan}}
  \newcommand{\apeking}{ \affiliation{Peking University, Beijing 100871, P.R.China}}
  \newcommand{\atum}{    \affiliation{Physik Department E12, Technische Universit\"at M\"unchen, 85748 Garching, Germany}}
  \newcommand{\alpc}{    \affiliation{LPC-Caen, ENSICAEN, Universit\'e de Caen, CNRS/IN2P3, 14050 Caen cedex, France}}
  \newcommand{\arikkyo}{ \affiliation{Department of Physics, Rikkyo University, Toshima, Tokyo 172-8501, Japan}}
  \newcommand{\auot}{    \affiliation{Department of Physics, University of Tokyo, Bunkyo, Tokyo 113-0033, Japan}}
  \newcommand{\atit}{    \affiliation{Department of Physics, Tokyo Institute of Technology, Meguro, Tokyo 152-8551, Japan}}
  \newcommand{\acns}{    \affiliation{Center for Nuclear Study, The University of Tokyo, RIKEN Campus, Wako, Saitama 351-0198, Japan}}
  \newcommand{\atus}{    \affiliation{Department of Physics, Tokyo University of Science, Noda, Chiba 278-8510, Japan}}
  \newcommand{\asaitama}{\affiliation{Department of Physics, Saitama University, Saitama 338-8570, Japan}}
  \newcommand{\agsi}{    \affiliation{GSI Helmholtzzentrum f\"ur Schwerionenforschung GmbH, 64291 Darmstadt, Germany}}

  \newcommand{\aem}{\email{scheit@ribf.riken.jp}}  

  \author{P.~Doornenbal}               \ariken             
  \author{H.~Scheit}             \aem  \ariken  \apeking   
  \author{N.~Aoi \kn{青井考}}            \ariken             
  \author{S.~Takeuchi \kn{武内聡}}       \ariken             
  \author{K.~Li \kn{李闊昂}}             \ariken  \apeking   
  \author{E.~Takeshita \kn{竹下英里}}     \ariken             
  \author{H.~Wang}                      \ariken \apeking   
  \author{H.~Baba \kn{馬場秀忠}}          \ariken            
  \author{S.~Deguchi}                   \atit              
  \author{N.~Fukuda \kn{福田直樹}}        \ariken            
  \author{H.~Geissel}                   \agsi              
  \author{R.~Gernh\"auser}              \atum              
  \author{J.~Gibelin}                   \alpc              
  \author{I.~Hachiuma}                  \asaitama          
  \author{Y.~Hara}                      \arikkyo           
  \author{C.~Hinke}                     \atum              
  \author{N.~Inabe \kn{稲辺尚人}}         \ariken            
  \author{K.~Itahashi \kn{板橋健太}}      \ariken            
  \author{S.~Itoh}                      \auot              
  \author{D.~Kameda \kn{亀田大輔}}        \ariken            
  \author{S.~Kanno \kn{菅野祥子}}         \ariken            
  \author{Y.~Kawada}                    \atit              
  \author{N.~Kobayashi}                 \atit              
  \author{Y.~Kondo \kn{近藤洋介}}         \ariken            
  \author{R.~Kr\"ucken}                 \atum              
  \author{T.~Kubo \kn{久保敏幸}}          \ariken            
  \author{T.~Kuboki}                    \asaitama          
  \author{K.~Kusaka \kn{日下健祐}}        \ariken            
  \author{M.~Lantz}                     \ariken            
  \author{S.~Michimasa}                 \acns              
  \author{T.~Motobayashi \kn{本林透}}     \ariken            
  \author{T.~Nakamura \kn{中村隆司}}      \atit              
  \author{T.~Nakao}                     \auot              
  \author{K.~Namihira}                  \asaitama          
  \author{S.~Nishimura \kn{西村俊二}}     \ariken            
  \author{T.~Ohnishi \kn{大西哲哉}}       \ariken            
  \author{M.~Ohtake \kn{大竹政雄}}        \ariken            
  \author{N.A.~Orr}                     \alpc              
  \author{H.~Otsu \kn{大津秀暁}}          \ariken            
  \author{K.~Ozeki \kn{大関和貴}}         \ariken            
  \author{Y.~Satou}                     \atit              
  \author{S.~Shimoura}                  \acns              
  \author{T.~Sumikama \kn{炭竃聡之}}      \atus              
  \author{M.~Takechi \kn{武智麻耶}}       \ariken            
  \author{H.~Takeda \kn{竹田浩之}}        \ariken            
  \author{K.~N.~Tanaka}                 \atit              
  \author{K.~Tanaka \kn{田中鐘信}}        \ariken            
  \author{Y.~Togano \kn{栂野泰宏}}        \ariken            
  \author{M.~Winkler}                   \agsi              
  \author{Y.~Yanagisawa \kn{柳澤善行}}    \ariken            
  \author{K.~Yoneda \kn{米田健一郎}}      \ariken            
  \author{A.~Yoshida \kn{吉田敦}}        \ariken            
  \author{K.~Yoshida \kn{吉田光一}}       \ariken            
  \author{H.~Sakurai \kn{櫻井博儀}}       \ariken            

  \date{\today}

  %
  %
  \begin{abstract}
    We report on the first spectroscopic study of the $N=22$ nucleus
    \ts{32}Ne
    at the newly completed RIKEN Radioactive Ion Beam Factory.
    A single \gammaray line with an energy of \neETwoPlus was
    observed in both inelastic scattering of a 226 MeV/$u$ \ts{32}Ne 
    beam on a Carbon target and proton removal from \ts{33}Na
    at 245 MeV/$u$.
    This transition is assigned to the de-excitation of the first $J^\pi = 2^+$ state 
    in \ts{32}Ne to the $0^+$ ground state.
    Interpreted through comparison with state-of-the-art shell model calculations, the low
    excitation energy demonstrates that the \ioi extends to at least $N=22$ for the Ne 
    isotopes.
  \end{abstract}

  \maketitle
%
%

One of the most fundamental concept in nuclear structure, 
as first introduced by Mayer and Jensen \cite{haxel:1949,mayer:1949},
is the notion of ``magic numbers''.
A nucleus with a certain number of protons and neutrons is said to be ``magic'' 
when large gaps occur in the single-particle (SP) energy spectra near the Fermi energy.
In this case residual interactions, which are weaker than the energy gap in the SP
spectrum, can only induce weak correlations and the nucleus exhibits
typical SP properties.
On the other hand, for smaller gaps or partially filled orbitals
the residual interactions can easily promote nucleons to SP states with a higher energy, 
giving rise to large correlations that are manifested in various collective phenomena.

While in the past magic neutron and proton numbers were considered static, i.e.\ independent of the
region in the nuclear chart being considered, it has become clear that
this is not the case and modifications of the standard shell ordering
occur in nuclei far from stability.
Currently considerable experimental and theoretical effort is being expended to uncover the
mechanisms driving these changes in shell structure \cite{sorlin:2008}.

Beyond ground-state binding energies, a variety of signatures exist to identify magic numbers.
One of the most direct is the reduced transition probability \beupl for even-even nuclei, 
which provides a measure of the correlations present in the wave functions.
Another key signature is the ratio of the energies of the first
$J^\pi=4^+$ and $2^+$ states.
For the most exotic  nuclei such information is often unavailable.
However, it has been shown that 
the  energy of the first $2^+$ state, \etwop, is a very good indicator of changes in 
nuclear structure \cite[Sec. 2-2b]{bohr-mottelson}.
More recently, and with a much larger data-set,
Cakirli and Casten \cite{cakirli:2008} have demonstrated that the \etwop alone provides 
a very strong signature of shell evolution.

The archetypical
example of very rapid changes in nuclear structure 
is the vanishing of 
the $N=20$ shell gap for the very neutron-rich Ne, Na and Mg isotopes, a region
which is now known as the \ioi \cite{warburton:1990}.
%
Soon after the pioneering work of Klapisch and Thibault \cite{klapisch:1969,thibault:1975} revealing
anomalies in the binding energies of the neutron-rich Na isotopes it was
suggested that the $\nu f_{7/2}$ orbitals actually intrude
into the $sd$ shell at $N=20$,
leading to a vanishing of the $N=20$ shell gap \cite{campi:1975}.
In a later seminal shell-model study of this region by Warburton \etal{}\cite{warburton:1990}
a true inversion of the orbitals was not found.
However, $\nu(sd)^{-2}(fp)^2$ ($2\hbar\omega$) intruder configurations
were predicted to become so low in energy that they form the
ground states for $Z=10$--12 and $N=20$--22, as subsequently confirmed by mass
measurements for neutron numers $N\le20$, 21 and 22 for the Ne, Na and Mg isotopes, respectively \cite{orr:1991}.
More recently, this behavior has been found to be a general phenomenon that should occur for most
standard shell closures far from stability and the mechanism 
behind this effect has been traced back
to the nucleon-nucleon tensor interaction by Otsuka and collaborators \cite{otsuka:2005PRL}.

The borders delineating the \ioi are by now rather well established on the high-$Z$ and low-$N$
sides \cite{sorlin:2008,gade:2008}.
For the Mg isotopes experiment confirms the dominance of the intruder configurations
for \ts{31-34,36}Mg, placing them inside the \ioi \cite{detraz:1979,motobayashi:1995,yoneda:2001,neyens:2005,gade:2007}
with a sharp transition from \ts{30}Mg, which is dominated by normal configurations \cite{niedermaier:2005PRL}.
For the Ne isotopes data are much more scarce but 
evidence available places \ts{30}Ne squarely inside the
\ioi with \ts{28,29}Ne at the boundary
\cite{orr:1991,iwasaki:2005,yanagisawa:2003,terry:2006}. 
Until now no spectroscopic data exists for the Ne isotopes with $N > 20$.

\begin{figure}
  \centering
  \includegraphics[width=8.3cm]{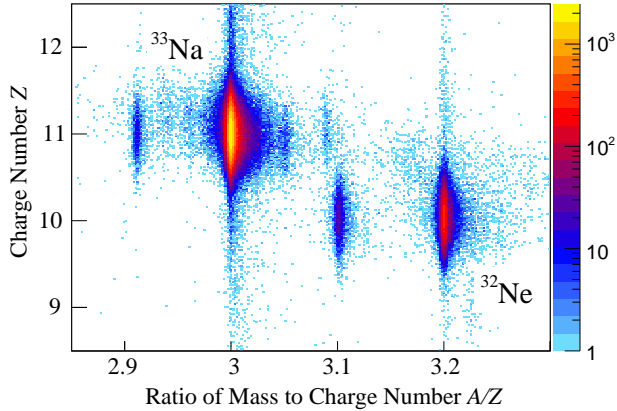}
  \caption{(color online) Particle identification before the secondary target.}
  \label{fig:pid-brips}
\end{figure}

%
%
Here, we report on the first spectroscopic study of the $N=22$ nucleus \ts{32}Ne.
%
%
The experiment was carried out at the recently commissioned Radioactive
Ion Beam Factory (RIBF) \cite{yano:2007} operated by the RIKEN Nishina Center and the Center for
Nuclear Study, University of Tokyo.
The secondary beams 
were produced by bombarding a 20 mm thick rotating Be
target \cite{yoshida:2008} with a \ts{48}Ca beam at 345 MeV/$u$ with an 
average intensity of $\sim 120$ pnA.
The projectile fragmentation products were analyzed and selected using the
standard magnetic rigidity, $B\rho$, selection method employing 
an achromatic Aluminum energy degrader of 15 mm median thickness located at the dispersive focus F1
\footnote{See \cite[Fig.\ 2]{kubo:2007} for the location of the focal points.}
of the first stage 
of the BigRIPS fragment separator \cite{kubo:2007,ohnishi:2008} (F0 to F2).
The momentum acceptance was $\pm3$\%.
The second stage of BigRIPS (F3 to F7) was used to identify the transmitted fragmentation
products using the $\Delta E$--$B\rho$--velocity method, where the energy loss
$\Delta E$ of the ions was measured in an ion chamber located at F7,
the $B\rho$ was determined from a position measurement (PPAC \cite{ohnishi:2008}) at the
dispersive focus F5 and the time-of-flight (TOF) was measured between
two thin plastic scintillators at F3 and F7 separated by a flight path
of 47 m.
The resulting particle identification (PID) is shown in Fig.\ \ref{fig:pid-brips},
where all isotopes are well separated---the resolution in $Z$ is 0.5 (FWHM) and the resolution in $A$ for the
Ne isotopes is 0.06 (FWHM).
The average secondary beam intensities were 6 \ts{32}Ne s\ts{-1} and 27 \ts{33}Na s\ts{-1}
in approximate agreement with the EPAX2 predictions \cite{summerer:2000}.
It should be noted that 
this corresponds to a gain in intensity of over two orders of
magnitude in comparison to the older RIPS facility.

The secondary beams were transported to the F8 focus where
a 2.54 g/cm\ts2 thick (natural) carbon target was mounted.
The mid-target energy of the $^{32}$Ne and \ts{33}Na beams was 226 MeV/$u$ and 245 MeV/$u$,
respectively. 
The energy loss in the target amounted to $\sim 13$\% of the incident beam energy in both cases.

The secondary target was surrounded by the DALI2 NaI(Tl) based \gammaray 
spectrometer \cite{takeuchi:2003} consisting of 
180 detectors covering
laboratory angles from 11$^\circ$ to 147$^\circ$.
The measured full energy peak efficiency was 
15 \% at 1.3 MeV, in agreement with GEANT4 simulations, 
and the resolution was 6\% (FWHM).

After the secondary reaction target the beam and reaction products entered the Zero Degree Spectrometer (ZDS)
\cite{mizoi:2005,kubo:2007} with angular and momentum acceptances of $\sim 80 \times 60$ mrad\ts2 and $\pm4$\%, respectively.
The ZDS provided the PID and the $B\rho$ was set
to that of elastically scattered \ts{32}Ne.
The overall transmission for elastically scattered \ts{32}Ne was $>90$\%.
Owing to the high acceptance of the ZDS the single-proton removal channel from
\ts{33}Na to \ts{32}Ne was observed simultaneously, albeit with a much lower transmission.
As before, the $\Delta E$--$B\rho$--TOF method was applied to 
identify the particles event-by-event, with a $B\rho$ measurement (PPAC) at the dispersive foci F9 and F10, 
a $\Delta E$ measurement at
the final focus F11 (ion chamber) and a TOF measurement between plastic scintillators 
mounted at F8 and F11 with a flight path of 37 m.
The resolutions in $Z$ and $A$ (for Ne) were 0.32 and 0.09 (FWHM), respectively.

%
%

\begin{figure}
  \centering
  \includegraphics[width=8.3cm]{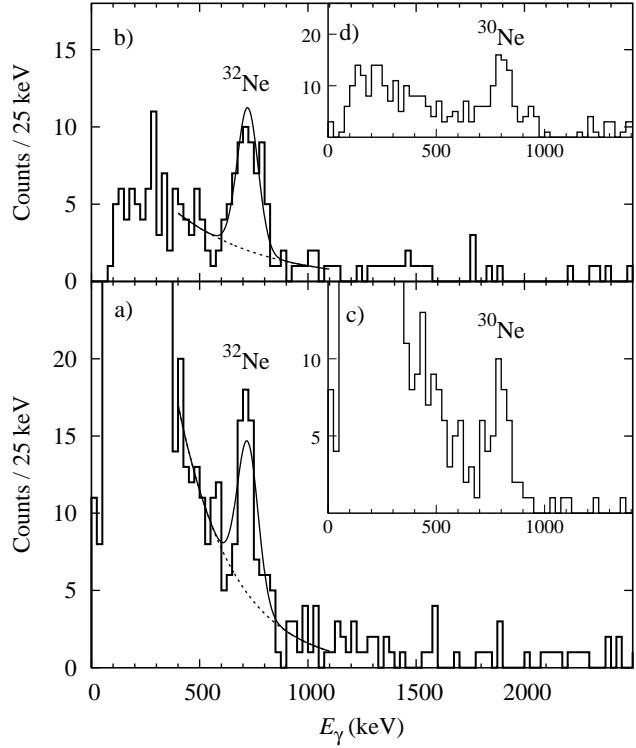}
  \caption{Doppler corrected \gammaray energy spectra in coincidence
    ($\pm5$ ns)
    with \ts{32}Ne (a,b) and \ts{30}Ne (c,d).
    Panel a) shows the results for inelastic scattering of \ts{32}Ne and 
    b) the result for proton removal from \ts{33}Na.
    The outcomes of the fitting procedure
    are shown by the solid (total) and dashed (background) curves.
    Here, both spectra were fitted simultaneously with
    the same peak position and peak width, but different peak areas and 
    background parameters.
    The inset panels c) and d) show the results for inelastic scattering of \ts{30}Ne and 
    for $p2n$ removal from \ts{33}Na, respectively, populating the first $2^+$ state
    in \ts{30}Ne.
  }
  \label{fig:spec}
\end{figure}

After a total measuring time of 8 hours a \gammaray transition, which we assign to the 
$2^+_1 \rightarrow 0^+_{gs}$ transition 
in $^{32}$Ne, could be clearly identified not only after inelastic excitation,
but also after single-proton removal from $^{33}$Na.
The Doppler corrected \gammaray energy spectra are shown in Fig.\
\ref{fig:spec} where the new transition can be clearly seen.
Fitting the sum spectrum with a Gaussian peak and an exponential background 
a transition energy of \neETwoPlus was derived, 
whereby the quoted uncertainty includes statistical (7 keV) and systematic (6 keV) contributions.
The latter are dominated by the unknown lifetime of the $2^+$ state resulting
in an uncertainty in the position and velocity of the \gammaray emitting particle.
The observed resolution is 15\% (FWHM).
Assuming no feeding transition (there is no evidence in either spectrum) 
the cross section for inelastic excitation on C of the first $2^+$
state was deduced to be 13(3) mb. 
Owing to the large uncertainty in the cross section and in the unknown optical model parameters 
at this beam energy, no meaningful deformation
parameter could be extracted from the cross section.

In order to study the effect of the lifetime of the excited state on the deduced
\gammaray energy a GEANT4 simulation was developed.
In addition to the full detector, target and beam line geometries, the simulation
(as well as the data analysis) took into account the lifetime of the
excited state, which is the principal contribution to the systematic uncertainty. 
The assumed lifetimes were varied in an interval from 
$0.5 \cdot \tau_R = 31$ ps to 
$2.0 \cdot \tau_R = 123$ ps, where
$\tau_R = 61$ ps was given by Raman's global systematics \cite[Eq. (11)]{raman:2001}.

As a check of our method we determined the \etwop of \ts{30}Ne produced after $p2n$ removal
from \ts{33}Na and after inelastic scattering of \ts{30}Ne---another BigRIPS setting was used 
for this measurement. 
Our result of 801(7) keV agrees well
with the literature value of 791(26) keV \cite{yanagisawa:2003}.
The corresponding spectra are displayed in the insets of Fig.\ \ref{fig:spec}. 

\begin{figure}
  \centering
  \includegraphics[width=8.3cm]{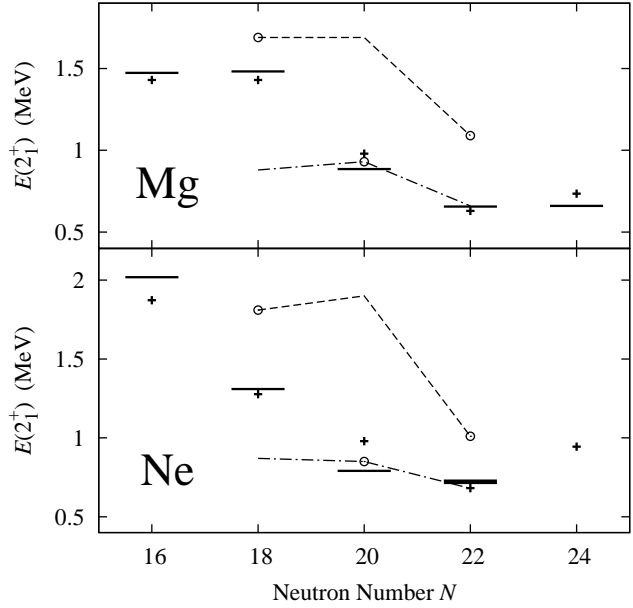}
  \caption{Comparison of experimental \etwop in neutron-rich Ne and Mg
    isotopes \cite{raman:2001,yanagisawa:2003,yoneda:2001,gade:2007}, indicated by 
    horizontal lines (the present work is shown in bold), with the shell
    model results of Utsuno \etal{}\cite{utsuno:1999} (+), and
    Caurier \etal{}\cite{caurier:1998,caurier:2001} for the normal ($\mathcal N$, dashed) and intruder
    ($\mathcal I$, dash-doted) configurations, respectively.
    Their prediction for the configuration ($\mathcal N$ or $\mathcal I$) 
    with the lowest energy $0^+$ state, i.e.\ the ground state, is marked
    by a circle.
    The \etwop are given relative to the $0^+$ state of the same configuration.
  }
  \label{fig:e-two-plus}
\end{figure}

%
%
%
We now turn to the interpretation of our results.
In Fig.\ \ref{fig:e-two-plus} the experimental \etwop values are shown 
as a function of neutron number. 
The very low \etwop at $N=20$ and 22 strongly suggest that there is
no $N=20$ shell gap and that \ts{32}Ne as well as \ts{30}Ne belong to the \ioi.
A very different behavior would be expected if $N=20$ was a good magic number. 
For instance, the \etwop as a function of neutron number for $14 \le Z \le 20$ 
exhibit a sharp peak at $N=20$ and 
where \etwop in excess of 2 MeV are observed.
Also, according to the ``$N_pN_n$''
scheme \cite{casten:1993}, in which the \etwop is correlated with the inverse of the
product of the number of valence neutrons $N_n$ and protons $N_p$,
a very different trend to that observed would be expected if $N=20$ were a good magic number.
Clearly this is not the case.

As long ago as 1990 Warburton \etal{}\cite{warburton:1990} predicted \ts{32}Ne to lie inside the \ioi, 
with essentially degenerate intruder ($2\hbar\omega$) and normal ($0\hbar\omega$) configurations.
Unfortunately no attempt was made to calculate the \etwop.
Even now only a few predictions for the \etwop of \ts{32}Ne 
exist \cite{utsuno:1999,siiskonen:1999,caurier:1998,caurier:2001,rodriguez-guzman:2003}.
Of particular interest are those of Utsuno \etal{}\cite{utsuno:1999} and
Caurier \etal{}\cite{caurier:2001} as they exhibit the best agreement with experimental data.
This is shown in Fig.\ \ref{fig:e-two-plus}, where
for completeness, the predictions and experimental data for the Mg
isotopes are also displayed.

The Monte Carlo shell model (MCSM) with the SDPF-M interaction of
Utsuno \etal allows
for a comprehensive theoretical exploration of nuclei in and around the \ioi
with unrestricted mixing of the $sd$ and $pf$ configurations \cite{utsuno:1999,otsuka:2001a}.
It provides an almost perfect description
of all \etwop in Fig.\ \ref{fig:e-two-plus}, including that of
\ts{32}Ne---the largest discrepancy occurs for \ts{30}Ne, but is less than 200 keV.
The MCSM calculations predict that the Ne and Mg isotopes with $N=20$ and 22 are strongly deformed
and dominated by intruder configurations \cite{utsuno:1999}.
Employing mean field calculations with the separable monopole interaction,
a similar conclusion with regard to the deformation was reached by Stevenson \etal{}\cite{stevenson:2002}
for \ts{32,34}Mg and \ts{32}Ne, but not \ts{30}Ne.
Utsuno \etal also predict the number of additional neutrons in the pf
shell with respect to normal filling.
Their result for the Ne and Mg isotopes with $N=20$ and
22, of $\langle n_{pf}\rangle \approx  n_{pf}^{\mbox{\tiny norm}} + 2$, agrees very well with the 
original predictions of Warburton \etal\ \ 
Deviations only occur for $N=18$ and $N=24$, where Utsuno \etal still anticipate
sizable intruder components in the ground-state wave functions, especially
for the Ne isotopes, while Warburton \etal do not.

Caurier \etal{}\cite{caurier:2001} performed separate
large-scale shell-model calculations
for the normal ($0\hbar\omega$) and intruder ($2\hbar\omega$) configurations, but
did not allow mixing of the two.
The results of the calculations are shown in
Fig.\ \ref{fig:e-two-plus}.
Contrary to Warburton \etal and Utsuno \etal they predict that only the
$N=20$ isotones have intruder ground states and reproduce the
\etwop in these cases.
While the intruder configuration result for \ts{32}Ne reproduces the experimental
value very well,
the normal configuration $0^+_1$ state is actually predicted to be the ground state
with the intruder $0^+_1$ state lying 1.5 MeV above it \cite{caurier:2001}.
The normal configuration $2^+_1$ state is predicted at 1 MeV
at variance with our observation.

Besides the expectation from the systematic trend of the \etwop energies,
where decreasing values with larger $N$ suggest increased
collectivity for \ts{32}Ne,
all model predictions consistent with the experimental \etwop---i.e.\ the 
predictions of Utsuno \etal{}\cite{utsuno:1999} and the intruder results of 
Caurier \etal{}\cite{caurier:2001}---show
\ts{32}Ne to be highly collective and therefore inside the \ioi with 
the ground state dominated by $2\hbar\omega$ intruder configurations.

In summary, we have reported on the first observation of a \gammaray line
($E_\gamma = $\neETwoPlus)
in in-beam \gammaray spectroscopy of \ts{32}Ne, which 
we assign to the $2^+_1\rightarrow 0^+_{gs}$ transition.
This measurement demonstrates that the \ioi extends to neutron number $N=22$ 
for the Ne isotopic chain.

It is interesting to note that thirty years after the first observation of an excited state in 
\ts{32}Mg \cite{detraz:1979}
experiment has progressed such that we can now report on
the same observation in the much more neutron-rich isobar \ts{32}Ne
from the first in-beam \gammaray spectroscopy experiment at the newly commissioned RIBF.
Developments in the near future should permit similar measurements to be
extended to other \ioi nuclei as well as Coulomb excitation studies.

\begin{acknowledgments}
  We would like to express our gratitude to the staff of the RIKEN Nishina Center
  for providing a stable and high intensity \ts{48}Ca beam.
  Furthermore, we wish to thank T. Ohtsubo and his group for allowing us to begin our measurements in 
  parallel with their own in ``Yakitori'' mode.
  P.D. and M.L. acknowledge the financial support of the Japan
  Society for the Promotion of Science.
  This work was in part supported by the DFG cluster of excellence {\it Origin and Structure of the Universe}.
\end{acknowledgments}

%
%
\bibliography{cb_a1}   

\end{document}